\documentclass[conference]{IEEEtran}
\IEEEoverridecommandlockouts
\usepackage{cite}
\usepackage{amsmath,amssymb,amsfonts}
\usepackage{algorithm}

\usepackage{textcomp}
\def\BibTeX{{\rm B\kern-.05em{\sc i\kern-.025em b}\kern-.08em
    T\kern-.1667em\lower.7ex\hbox{E}\kern-.125emX}}

\usepackage[pdftex]{graphicx}

\usepackage{amsmath}

\usepackage[noend]{algpseudocode}

\usepackage{array}

\ifCLASSOPTIONcompsoc
 \usepackage[caption=false,font=normalsize,labelfont=sf,textfont=sf]{subfig}
\else
 \usepackage[caption=false,font=footnotesize]{subfig}
\fi

\usepackage{stfloats}
\usepackage{url}

\usepackage{amssymb} 
\usepackage{multirow}
\usepackage[bookmarks=false]{hyperref}  
\usepackage[usenames,dvipsnames,svgnames]{xcolor}
\usepackage{colortbl}
\usepackage{hhline}
\usepackage{threeparttable} 
\usepackage{makecell}

\usepackage{pgfplots}
\pgfplotsset{compat=1.18}

\usepackage{mathtools}
\DeclarePairedDelimiter{\ceil}{\lceil}{\rceil}

\makeatletter
\let\old@ps@headings\ps@headings
\let\old@ps@IEEEtitlepagestyle\ps@IEEEtitlepagestyle
\def\confheader#1{%
\def\ps@IEEEtitlepagestyle{%
\old@ps@IEEEtitlepagestyle%
\def\@oddhead{\strut\hfill#1\hfill\strut}%
\def\@evenhead{\strut\hfill#1\hfill\strut}%
}%
\ps@headings%
}
\makeatother

\confheader{%
\textit{To appear in the International Conference on Field Programmable Technology (FPT'23), Yokohama, Japan, 11-14 December 2023.}
}

\usepackage[pscoord]{eso-pic}
\newcommand{\placetextbox}[3]{
\setbox0=\hbox{#3}
\AddToShipoutPictureFG*{ \put(\LenToUnit{#1\paperwidth},\LenToUnit{#2\paperheight}){\vtop{{\null}\makebox[0pt][c]{#3}}}
}
}
\placetextbox{.5}{0.055}{\small \copyright 2023 IEEE. Personal use of this material is permitted. Permission from IEEE must be obtained for all other uses, in any current or}
\placetextbox{.5}{0.045}{\small future media, including reprinting/republishing this material for advertising or promotional purposes, creating new collective works,}
\placetextbox{.5}{0.035}{\small for resale or redistribution to servers or lists, or reuse of any copyrighted component of this work in other works.}

\begin{document}

\title{A High-Frequency Load-Store Queue with Speculative Allocations for High-Level Synthesis}

\author{\IEEEauthorblockN{Robert Szafarczyk, Syed Waqar Nabi and Wim Vanderbauwhede}
\IEEEauthorblockA{School of Computing Science\\
University of Glasgow, UK
\\ Email: \{robert.szafarczyk, syed.nabi, wim.vanderbauwhede\}@glasgow.ac.uk}}

\maketitle

\begin{abstract}

Dynamically scheduled high-level synthesis (HLS) enables the use of load-store queues (LSQs) which can disambiguate data hazards at circuit runtime, increasing throughput in codes with unpredictable memory accesses.
However, the increased throughput comes at the price of lower clock frequency and higher resource usage compared to statically scheduled circuits without LSQs.
The lower frequency often nullifies any throughput improvements over static scheduling, while the resource usage becomes prohibitively expensive with large queue sizes.
This paper presents a method for achieving dynamically scheduled memory operations in HLS without significant clock period and resource usage increase.
We present a novel LSQ based on shift-registers enabled by the opportunity to specialize queue sizes to a target code in HLS.
We show a method to speculatively allocate addresses to our LSQ, significantly increasing pipeline parallelism in codes that could not benefit from an LSQ before.
In stark contrast to traditional load value speculation, we do not require pipeline replays and have no overhead on misspeculation.
On a set of benchmarks with data hazards, our approach achieves an average speedup of 11$\times$ against static HLS and 5$\times$ against dynamic HLS that uses a state of the art LSQ from previous work.
Our LSQ also uses several times fewer resources, scaling to queues with hundreds of entries, and supports both on-chip and off-chip memory.

\end{abstract}

\begin{IEEEkeywords}
high-level synthesis, load-store queue, compiler speculation, dynamic scheduling
\end{IEEEkeywords}

\section{Introduction}

High-level synthesis (HLS) tools transform high-level software code into a custom architecture that can be synthesized on an FPGA.
Such architectures have the potential to achieve higher performance and energy efficiency than general-purpose CPUs and GPUs \cite{efficient_communication_analysis_adler}.
A major obstacle to the wider adoption of FPGA acceleration remains their programmability.
HLS tools have lowered the barrier of entry for FPGA programmers dramatically when compared to using hardware description languages, but they still impose a specific structure on the input code, which is not intuitive to software programmers.
Our goal is to increase the quality of HLS by shifting the burden of structuring code for a spatial architecture from the programmer to the compiler.

Loop pipelining is a critical step in HLS.
It is the process of starting new loop iterations while previous iterations have not yet finished, allowing to achieve higher throughput with the same amount of compute resources.
The number of cycles between the start of two subsequent iterations is called the Initiation Interval (II).
A loop with a constant II, $N$ iterations, and a latency of $L$ will execute in $L + (N - 1) \times II$ cycles, which for $N \gg L$ can be approximated as $N \times II$.
Thus, a low loop II is crucial to achieving good performance in HLS.

Most HLS tools use modulo scheduling to perform loop pipelining \cite{modulo_sched, modulo_sched_canis, modulo_scheduling_koch} (such tools are often called static HLS).
Modulo scheduling maps operations for a single loop iteration to discrete clock cycles at compile time and then repeats this schedule for all loop iterations.
One of the first steps in modulo scheduling is determining the minimum number of cycles between the start of subsequent loop iterations, while honoring data dependencies across iterations.
Such data dependencies form recurrences in the Data Dependence Graph (DDG) of the input code. 
Modulo scheduling finds the maximum recurrence-constrained II across all recurrences for a given loop:
$$recII = max_i  \lceil delay_i/distance_i \rceil,$$
where $delay$ is the number of cycles needed to traverse the recurrence path, and $distance$ is the number of iterations between the definition of a recurrence value and its use.

\begin{figure}[t!]
\centering
\subfloat[Motivating source code with a data hazard.]{\includegraphics[width=0.38\textwidth]{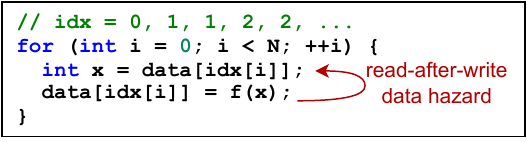}\label{fig:MotivatingExampleCode}}%
\vfill%
\subfloat[A static schedule: a new iteration started every 3 cycles for all iterations.]{\includegraphics[width=0.48\textwidth]{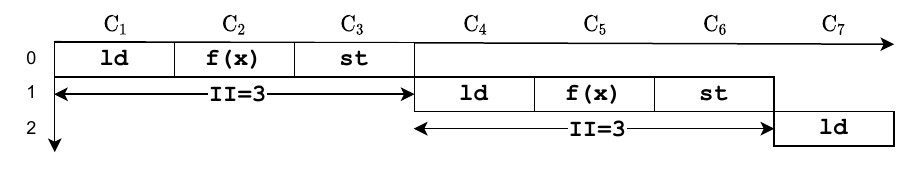}\label{fig:MotivatingExampleStaticSchedule}}%
\vfill%
\subfloat[An ideal schedule: a new iteration started every 1.5 cycles on average.]{\includegraphics[width=0.48\textwidth]{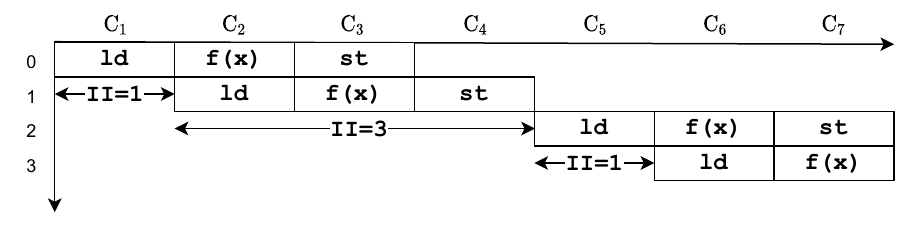}\label{fig:MotivatingExampleIdealSchedule}}%

\caption{A motivating example of code with a data hazard. Current static HLS tools need to create a worst case schedule at compile time (b). HLS with dynamically scheduled memory operations can achieve the schedule in (c).}

\label{fig:MotivatingExample}
\end{figure}

Static HLS tools rely on an accurate memory dependency analysis to discover the dependence $distance$ of a DDG recurrence through memory.
Memory dependency analysis from software compilers, such as the polyhedral model, are directly applicable in this case \cite{poly_hls, liu_pipelining_polly, dependence_distance_hls}. 
However, there is a large class of codes where the calculation of the dependence distance is fundamentally impossible due to limited compile time information.
Take the code in fig. \ref{fig:MotivatingExample} as an example.
The code contains data-dependent memory reads and writes that form a recurrence in the DDG.
For such codes, the dependence distance cannot be obtained and has to be conservatively set to one, i.e. it is assumed that every iteration needs to wait for all previous iterations to finish, eliminating any possibility for loop pipelining as seen in fig. \ref{fig:MotivatingExampleStaticSchedule}.





An alternative approach to achieving loop pipelining is to use dynamic scheduling.
Dynamic HLS uses dataflow scheduling to trigger the execution of operations at runtime based on the availability of data, rather than a static compile-time schedule \cite{josipovic_dynamatic_2022, Townsend_Kim_Edwards_2017}.
Dynamic scheduling enables the use of load-store queue (LSQ) structures that allow for dynamically scheduled out-of-order loads that are essential for pipelining codes with unpredictable memory accesses \cite{Josipovic_Brisk_Ienne_2017}.
For our example code from fig. \ref{fig:MotivatingExampleCode}, dynamic HLS with an LSQ can achieve the ideal schedule in fig. \ref{fig:MotivatingExampleIdealSchedule}.
However, dynamic HLS incurs non-trivial resource and critical path overheads, often nullifying any throughout advantage over static HLS \cite{josipovic_dynamatic_2022, dynamic_inter_block_cheng}.
Recent work has shown the possibility of combining static and dynamic scheduling to achieve the high throughput of dynamic scheduling with the low critical path of static HLS \cite{dass, szafarczyk_fpl23}.
However, whenever an LSQ is needed by the dynamic part of such a combined circuit, the high critical path and area overheads return.

Thus, to unleash the full potential of circuits combining dynamic and static scheduling, there is a clear need for an LSQ with a faster critical path, lower area overhead, and better scalability than previous work.
We make the following contributions toward this goal:
\begin{itemize}
    \item A novel load-store queue (LSQ) design for HLS with shift-register based queues enabled by the opportunity to specialize queue sizes to a target code in HLS (sec. \ref{sec:lsqDesign}).
    We show how the decoupled access/execute architecture model can be applied to address generation to enable the use of an LSQ in static HLS (sec. \ref{sec:compiler}). 
    \item An extension to our LSQ and compiler that enables \textit{speculative address allocations} -- a compiler speculation algorithm applied to LSQ address allocations in HLS that doesn't require replays and has no misspeculation penalty (sec. \ref{sec:speculation} and \ref{sec:speculationGeneral}).
    \item An evaluation of our work against static HLS (Vivado HLS, Intel HLS), and against dynamic HLS. 
    We show that our work achieves both better speedup and lower area overhead compared to a dynamic HLS compiler with a state of the art LSQ \cite{josipovic_dynamatic_2022}. 
    We demonstrate that our LSQ supports larger queues (tab. \ref{table:scalabilityTableQueue}), and that our speculative LSQ address allocations enable to accelerate a broader range of codes (sec. \ref{sec:evaluation}).
    We show that our LSQ can be used equally well to protect on-chip and off-chip memory.
\end{itemize}

\section{Background \& Related Work} \label{sec:background}


\subsection{Dynamically Scheduled High-Level Synthesis} \label{sec:backgroundDynSched}

Dynamically scheduled circuits rely on the theory of latency-insensitive design formalized by Carloni \textit{et al.} \cite{Carloni_McMillan_Sangiovanni_Vincentelli_2001} and simplified by Cortadella \textit{et al.} for synchronous circuits \cite{Cortadella_Kishinevsky_Grundmann}.
In latency-insensitive designs, the communication between modules is decoupled from their cycle behavior, allowing for dataflow scheduling of compute circuits \cite{Fleming13_lic_thesis, Kam_Kishinevsky_Cortadella_micro_pipelining, huang2013elastic, elastic_flow, josipovic_dynamatic_2022, Townsend_Kim_Edwards_2017}.
By using more resources to defer scheduling to runtime, dataflow circuits can achieve perfect throughput on codes with unpredictable inter-iteration dependencies.
The disadvantage of dataflow circuits mapped to FPGA technology is firstly their significantly higher critical path, and secondly their higher area usage.
The higher area usage is often acceptable, but a higher critical path means that the final design synthesized on FPGA hardware is not able to achieve the frequencies achievable by static HLS. 
The critical path increase is due to using LSQs, and due to using buffers with a zero cycle write-to-read latency (called transparent buffers in Dynamatic \cite{josipovic_dynamatic_2022}) where static HLS can use a wire with a finite state machine controller.

\subsection{Combining Static and Dynamic Scheduling}

Cheng \textit{et al.} extended Dynamatic with the DASS methodology (Dynamic and Static Scheduling) \cite{dass, cheng_finding_static}, which identifies static islands in an otherwise dynamically scheduled circuit.
This improves the resource usage of the final circuits but the critical path stays often the same.

Szafarczyk \textit{et al.} extended modulo-scheduled HLS tools with support for selective dynamic scheduling by breaking up the DDG of an input code into multiple modulo-scheduling instances based on compiler analysis that determines where dynamic scheduling is beneficial \cite{szafarczyk_fpl23}.
The separate modulo scheduling instances communicate via latency-insensitive channels -- a construct available in most static HLS tools.
Their approach achieves virtually the same frequency as static HLS on codes that don't require an LSQ.
If an LSQ is required, the frequency of their approach matches that of Dynamatic. 
Since most codes amenable to dynamic scheduling do have unpredictable memory accesses that do require an LSQ, their approach is of limited value without an LSQ that can provide a low critical path.
In this work, we combine their scheduling methodology with a novel LSQ design that is able to achieve such low critical paths.

\subsection{Runtime Memory Disambiguation in HLS}

To avoid pipeline stalls due to unpredictable memory accesses, a circuit can use additional logic, such as load-store queues (LSQs), to handle memory accesses at runtime \cite{mem_disamb_approaches}.
If proven safe to do so, the logic should allow loads from later loop iterations to be executed without waiting for stores from earlier iterations to commit.
There are two main approaches to enable such out-of-order loads: address-based approaches compare addresses of loads and stores; value-based approaches speculatively execute loads and replay the datapath on misspeculation.

\textit{Value-based disambiguation:}
Thielmann \textit{et al.} investigated the use of load speculation in reconfigurable hardware \cite{Thielmann_Load_Speculation}.
In their framework, if a speculated load value turned out to be incorrect, then only the computation depending on the load had to be repeated, not the whole pipeline.
Nonetheless, codes with loop-carried dependencies, which are the focus of our work, had a high misprediction penalty that was a problem.
Dai \textit{et al.} \cite{dai_speculative_data_hazards} also used value speculation to enable pipelining of loops with irregular memory accesses.
They proposed a source-to-source transformation that replaces hazardous accesses with virtualized accesses to an independent array.
These independent array accesses are then handled by a custom Hazard Resolution Unit which speculatively executes loads, performs store-load forwarding, and sends misprediction signals to the datapath.
Misprediction triggers a squash and replay action, which adds overhead.
The benefit of value-based disambiguation is that it can pipeline loops where the store operation is control-dependent on a load \cite{Thielmann_Load_Speculation}.
The disadvantage is that squash-and-replay is prohibitively expensive.
Budiu \textit{el al.}, who developed one of the first dynamically scheduled HLS compiler \cite{budiu_thesis, Venkataramani_Budiu_Chelcea_Goldstein_2004}, noted that ``implementing a generic prediction scheme (be it branch prediction or value prediction) in a dataflow model is hindered by the difficulty of building a mechanism for squashing the computation on the wrong paths" \cite{Budiu_Artigas_Goldstein_2005}. 
We address this fundamental issue by proposing an LSQ and a compiler transformation that can disambiguate memory accesses on speculated paths with no requirement for squash-and-replays, i.e. with no misspeculation cost.

\textit{Address-based memory disambiguation} compares the addresses of loads and stores out-of-order with the actual load/store operations, allowing non-conflicting loads to execute even if earlier stores have not yet committed.  
Such functionality is most often implemented as an LSQ.
Most LSQs aimed at HLS have a similar operating principle as LSQs used in out-of-order CPUs \cite{mem_disamb_approaches}.
For example, the Dynamatic LSQ \cite{Josipovic_Brisk_Ienne_2017} has a single store queue buffer which holds stores in-flight to memory, together with metadata needed to recover program order.
Dependent loads check this structure for aliasing using the memory address and other metadata, deciding if a load is safe to perform, if a store value can be forwarded, or if the load has to wait.
It is this single-cycle Content-Addressable Memory (CAM) access that maps poorly to FPGA technology, resulting in a high critical path and area usage \cite{liu_lsq_sizing}.

Our LSQ design is fundamentally different.
We recognize that LSQs for HLS don't have to be as general as CPU LSQs.
We propose to break up the single store queue CAM into two separate shift-register based queues, one holding just store address allocations and the other store commits.
Compiler analysis allows us to size the shift-registers exactly.
Instead of the single-cycle CAM access in Dynamatic, we spread our memory disambiguation checks into multiple pipeline stages for an improved critical path and resource usage.
Another major difference is our support for speculative address allocations, enabled by having separate store allocation and commit queues.
Our LSQ approach can be seen as a generalization of shift-registers based approaches to pipelining of loops with statically analyzable dependency distances \cite{dac_runtime_dep_check_with_shiftreg}, e.g., sparse matrix-vector multiply accelerators \cite{spmv_fccm23}.

The central question in LSQ designs for spatial computing is how to recover program order of memory requests without a program counter.
Josipović \textit{et al.} proposed to allocate LSQ addresses from a single basic block in parallel and sequentialize the execution of basic blocks \cite{Josipovic_Brisk_Ienne_2017}.
Memory operations within a single basic block can be disambiguated statically, while the semantics of their dataflow circuits guaranteed the sequential execution of basic blocks in program order.
Our LSQ doesn't rely on the sequential execution of basic blocks.
Instead, we recover program order by tagging each memory request with a unique integer representing the state of memory at that time.
Our tags are similar to the work by Elakhras \textit{et al.} \cite{straight_to_the_queue} who addressed the sequentialized block allocation problem of the Dynamatic LSQ by introducing virtual data dependencies between blocks with LSQ accesses. 
However, in addition to ordering the allocation of addresses, we also use the actual tag values for disambiguation inside the LSQ.

\section{The Memory Disambiguation Problem} \label{sec:ProblemStatement}

We define an LSQ allocation as an $(address, tag)$ tuple.
The tag is an integer indicating the state of memory expected by the allocation.
We define memory states as a sequence $\sigma = \{0, 1, 2, ...\}$, where each $i \in \sigma$ corresponds to the memory state of the original sequential program after the $i$-th store, with the state at $i=0$ representing the initial memory state.

The inputs to our LSQ are: a sequence of load allocations; a sequence of store allocations; a sequence of store values where each $stValue_{i}$ corresponds to the $stAllocation_{i}$.
We require that store allocations and store values arrive in program order.
The LSQ outputs a sequence of load values, which correspond to the sequence of previously made load requests.

The tag of a load allocation indicates which memory state is expected by the load; the tag of a store allocation represents the new memory state after the store.
Given any pair of $ldAllocation_{i}$ and $stAllocation_k$, if the two conditions hold:
\begin{equation} \label{eq:memDisambiguation}
\begin{split}
ldAllocation_{i}.address &= stAllocation_k.address ,\\
ldAllocation_{i}.tag &\ge stAllocation_k.tag,
\end{split}
\end{equation}
then $ldAllocation_{i}$ cannot be served before observing the side-effect of $stAllocation_{k}$.

Finally, we define a store commit as an $(address, value)$ tuple.
Our LSQ holds a sequence of store commits internally, representing values in-flight to memory.
Store commits can be used to forward stored values directly to aliasing loads.
Note the omission of program ordering information from the store commits.
In previous LSQs, in the case when a load aliases multiple store commits, the forwarding logic had to pick the youngest store commit.
In our case, this would require adding a $tag$ field to the store commit tuple, and finding a store commit with the maximum $tag$ value.
We avoid the need for this logic by keeping store commits ordered, and by ensuring that the store commits don't contain stores that in program order come after a load that has not yet been served.


\section{Load-Store Queue Design} \label{sec:lsqDesign}








\begin{figure*}[t!]
\centering

\includegraphics[width=0.95\textwidth]{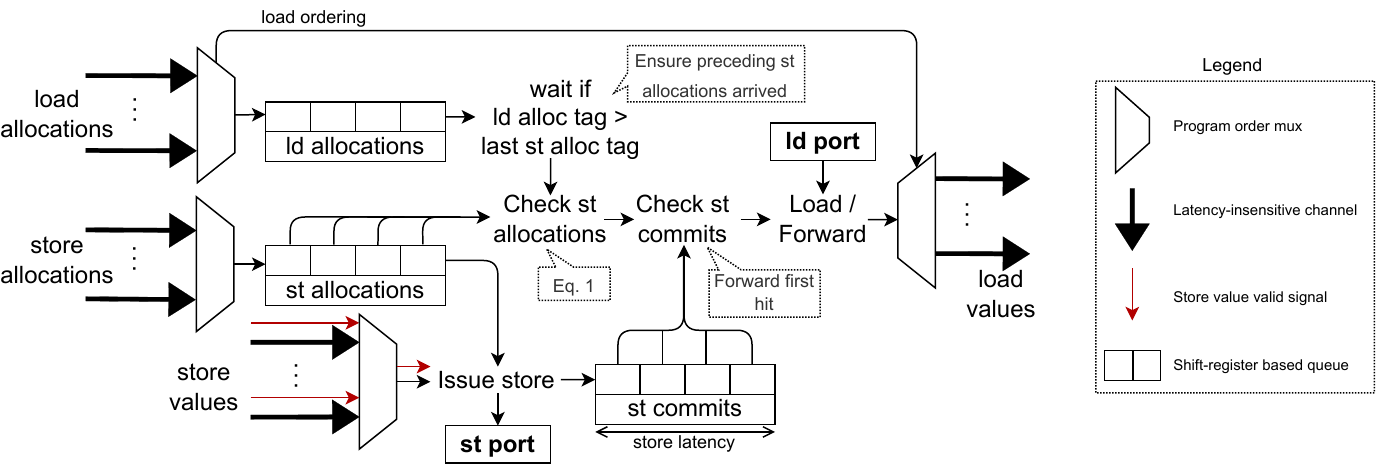}

\caption{Our shift-register based load-store queue design. Load allocations are checked for aliasing over multiple pipeline stages. Store values carry an optional $valid$ bit, allowing the LSQ to drop store allocations corresponding to invalid store values, and thus enabling speculative address allocations.}

\label{fig:LSQDesign}
\end{figure*}

We now present the design of our load-store queue (LSQ).
We show how loads and stores are executed, and how we support speculative address allocations.
We also discuss how our LSQ can protect both on-chip and off-chip memory.

\subsection{Load-Store Queue Overview}

Fig. \ref{fig:LSQDesign} shows an overview of our LSQ protecting a memory with one load and one store port.
Load/store allocation queues and the store commit queue are implemented as shift registers in FIFO order.
The store queue is broken up into two separate queues: one for allocations and one for commits.
Store-forwarding is the only time-critical logic in an LSQ, and decoupling it allows the rest of the LSQ to be pipelined.
Such decompositions have been proposed before for CPU LSQs \cite{decompose_lsq}.
We implement store-forwarding using a store commit queue, which holds stores for the duration between store issue and memory commit -- its size must be equal to the maximum store latency, which is program-specific.

Our LSQ accepts one load allocation per cycle for every available load port to memory.
Multiple load allocation sequences can be served in parallel as long as the number of sequences is not greater than the number of load ports.
If there are more load allocation sequences than available load ports to memory, then the sequences are multiplexed according to program order (as is the case in fig. \ref{fig:LSQDesign}).
Multiple store allocation sequences, and their corresponding store value sequences, are always multiplexed in program order, regardless of the amount of memory store ports available.
This restriction protects against write-after-write hazards by construction.

\subsection{Load and Store Execution} \label{sec:LSQLoadStoreExec}

\textit{Load execution:}
A given $ldAllocation_i$ at the head of the load allocation queue compares its tag to the latest accepted store allocation tag, and waits if its tag is higher.
This tag check ensures that all store allocations coming before $ldAllocation_i$ in program order have arrived to the LSQ.
Next, $ldAllocation_i$ checks all store allocations in the store allocation queue for conflicts using eq. \ref{eq:memDisambiguation}.
If there are no conflicts within the store allocation queue, then the store commit queue is checked next.
At this point, the store commit queue is guaranteed to hold only stores that come before $ldAllocation_i$ in program order.
In the commit queue, we check from the youngest to the oldest store and forward the first (i.e. youngest) value that matches the $ldAllocation_i$ address.
If there is no hit in the commit queue, then we can safely load the value from memory and return it via a non-blocking latency-insensitive channel to the datapath.

\textit{Store execution:}
A given $stAllocation_j$ at the head of the store allocation queue waits for its corresponding store value to arrive.
On the arrival of the awaited store value, a store is immediately issued to memory and a $stCommit_j$, holding the store address and store value, is shifted into the store commit queue.
The corresponding $stAllocation_j$ is shifted away from the store allocation queue.
A store can only be in the store allocation or store commit stage, but never both.
The store commit queue is sized such that it holds on to the store value until it is guaranteed to have been committed to memory.

\textit{Speculation support:}
Our LSQ can support \textit{speculative store allocations} by extending each store value with a valid bit.
Valid store values are handled without change.
Invalid store values are not stored to memory and are not shifted into the store commit queue.
Invalid store values still cause the corresponding store allocation to be shifted away.
This mechanism allows to speculatively allocate store addresses to the LSQ with no requirement for replays because \textit{a misspeculated store allocation is never actually committed}.
Sec. \ref{sec:speculationGeneral} shows how the compiler creates speculative allocations.

\subsection{Scalability to Off-Chip Memories}

Our LSQ design can be used to protect both on-chip and off-chip memory from data hazards.
Our LSQ can exploit multiple load ports in parallel.
Multiple store ports cannot be exploited by our design -- to protect write-after-write hazards, we multiplex multiple store sequences onto one store port.

To support multi-cycle memory we grow the size of the store commit queue to cover the maximum store latency.
To avoid stalls in the LSQ when issuing a multi-cycle variable-latency memory operation, we decouple the load and store ports from the LSQ pipeline and connect them using latency-insensitive buffers with a deterministic write-to-read latency.
To preserve the correctness of memory disambiguation, we grow the store commit queue by this added latency.

\section{Compiler Integration} \label{sec:compiler}

In this section, we first show how an HLS compiler can use our LSQ.
Then, we describe how to enable dynamically scheduled out-of-order loads in static HLS.
Next, we present a compiler algorithm for introducing speculative address allocations to our LSQ.
Finally, we show how parts of the LSQ can be specialized based on the target code.

We use existing compiler analysis' to find memory base addresses with data hazards \cite{Optimizing_Compilers}.
Each selected base address uses its own LSQ.
All memory operations using a selected base address are be transformed into read/writes from/to latency-insensitive channels connected to an LSQ.
The channels to an LSQ can be reused across basic blocks if they are guaranteed not to execute in the same clock cycle, similar to how FPGA block RAM ports can be shared.

Our LSQ design uses integer tags to recover program order of memory operations. 
Each address generating unit has a tag corresponding to a single LSQ, initially set to zero. 
Store allocations increment the tag before using it; load allocations use the tag directly.
This creates a data dependency between a store allocation and any other LSQ allocation following that store allocation in program order, thus ensuring the correct order of the store allocation sequence.

\subsection{Dynamically Scheduled Memory in Static HLS} \label{sec:decoupledAddressGen}

\begin{figure}[t]
\centering
\includegraphics[width=0.43\textwidth]{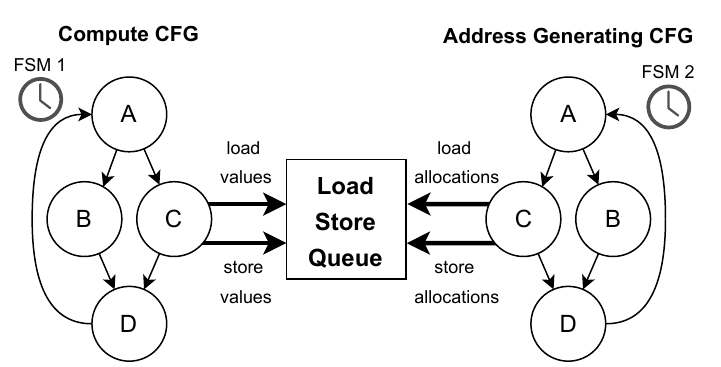}\label{fig:introCFGc}

\caption{The generation of addresses is decoupled into a separate modulo-scheduling instance, capable of generating load-store queue address allocations out-of-order w.r.t. the actual memory accesses in the compute loop.}

\label{fig:AddressDecoupling}
\end{figure}

Altough our LSQ can be used by any HLS tool, in this paper we assume it is used by a statically scheduled HLS compiler.
This subsection describes the transformation needed in static HLS to enable the efficient use of our LSQ and can be omitted if the LSQ is used in fully dynamic HLS.

The throughput of circuits using an LSQ depends on the number of addresses that can be disambiguated ahead of their actual memory operation execution -- call this the \textit{out-of-order address window}.
In a statically scheduled pipeline, the out-of-order address window can be at most one -- address generation and memory access proceed in lockstep.
In dataflow circuits, the generation of memory addresses is naturally decoupled from the memory operation and allows for much larger out-of-order address windows.
To achieve the same effect in static HLS, we follow a method from our previous work \cite{szafarczyk_fpl23} to decouple the generation of memory addresses into a separate static pipeline, similar to the principle in decoupled access/execute architectures \cite{decoupled_access_exec, decoupling_address_for_lat, decoupled_memory_prefetching}.
Fig. \ref{fig:AddressDecoupling} illustrates the resulting communication pattern.
The address generating unit will contain only address generating instructions and will run ahead w.r.t. the compute unit, increasing the out-of-order address window in our LSQ.

\subsection{Loss of Address Decoupling} \label{sec:lossOfDecoupling}

In some cases, address generation decoupling cannot result in the run-ahead of address allocations.
Such ``loss of decoupling'' \cite{decoupling_address_for_lat} arises when the address generation for a given base address depends on values loaded from the same base address, i.e. a load value from an array is used to generate a load/store address to the same array. 
Formally, given a set of address generating instructions $G$ for a given base address, and a set of memory access instructions $A$ using addresses generated by instructions in $G$, there is a loss of decoupling if:
\begin{itemize}
    \item[] $\exists i \in G$, such that $i$ depends on an instruction $j \in A$, i.e. there is a path from $i$ to $j$ in the DDG.
\end{itemize}
We do not perform address decoupling in such cases, because the address allocations and their memory operations need to be in effect synchronized.
This is not a drawback of using static HLS since a fully dynamically scheduled circuit would also synchronize the two sequences.

Note that our loss of decoupling definition is more relaxed than previous work.
We only consider direct data dependencies, ignoring control dependencies.
We next show how we use speculation to maintain decoupling of address generation in cases where a memory operation is control dependent, such that the control decision itself depends on a loaded value from the LSQ.
Our approach allows us to maintain a high out-of-order address window, even in cases where a fully dynamic HLS compiler would suffer a loss of decoupling.

\begin{figure}[t!]
\centering

\subfloat[Maximal Matching code and its control-flow graph (CFG).]{\includegraphics[width=0.44\textwidth]{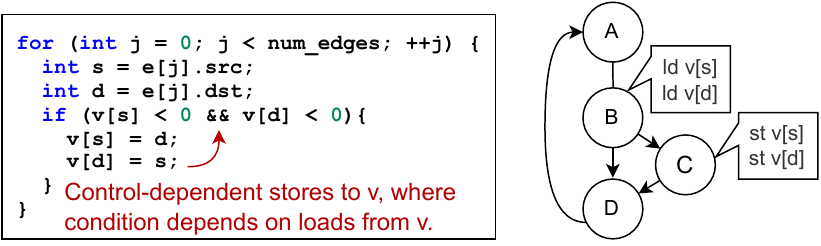}\label{fig:mmOriginal}}%
\vfill%
\subfloat[Our transformation: speculative address allocations in the address generation loop (left), and invalidated store value writes on misspeculation (right).]{\includegraphics[width=0.46\textwidth]{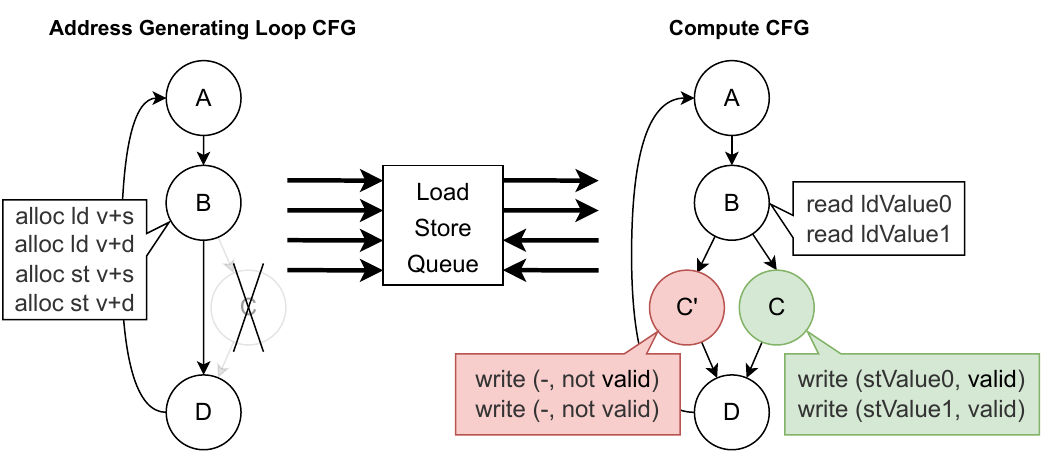}\label{fig:mmDecoupled}}%

\caption{Speculative store address allocations in the maximal matching code.}

\label{fig:SpeculativeStoreAlloc}
\end{figure}

\subsection{Intuition for Speculative Address Allocations} \label{sec:speculation}

A memory operation using a given base address can be control-dependent on a branch condition that itself is data-dependent on a value loaded from the same base address.
Consider the code in fig. \ref{fig:mmOriginal} as an example.
Here, the execution of the stores to \texttt{v} is control dependent on the if-condition which itself uses values loaded from \texttt{v}.
Under the execution model of both dynamic HLS \cite{josipovic_dynamatic_2022} and our decoupled address architecture (sec. \ref{sec:decoupledAddressGen}), there is no possibility for out-of-order address allocations in this code.
We propose the concept of \textit{speculative address allocations} to relax this restriction.

Consider the code in fig. \ref{fig:mmOriginal} again.
Although the store execution is control-dependent, the store addresses have no data dependency on values loaded from \texttt{v}.
We can hoist the address instructions out of the if-condition in the address generating CFG (fig. \ref{fig:mmDecoupled}).
As a result, store address allocations will be produced without having to evaluate the if-condition.

Addresses allocated to the LSQ, but later not used, are said to be \textit{misspeculated}.
Misspeculations are handled in the compute CFG by inserting invalid LSQ store value writes on CFG paths containing misspeculations (e.g. basic block $C'$ in fig. \ref{fig:mmDecoupled}).
An invalid LSQ store has the $valid$ bit set to 0 and will result in the deallocation of the misspeculated address allocation in the LSQ (sec. \ref{sec:LSQLoadStoreExec} describes the LSQ support).
Handling misspeculated loads is trivial, since a load doesn't have side effects and the loaded value can simply be discarded.

This compiler speculation approach can achieve a high degree of out-of-order loads such as in fig. \ref{fig:mmOriginal}, without having to suffer the cost of expensive misspeculation replays common in load-value-based speculation approaches.

\subsection{Compiler Generated Speculative Address Allocations} \label{sec:speculationGeneral}

\begin{figure}[t!]
\centering

\subfloat[Iterative hoisting of speculative address allocations (green blocks) to their special-control dependency source block.]{\includegraphics[width=0.24\textwidth]{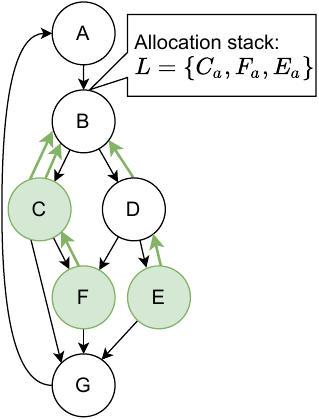}\label{fig:generalSpeculationAddress}}%
\hfill%
\subfloat[Insertion of poison basic blocks (red blocks) with invalidating loads/stores to deque misspeculated address allocations.]{\includegraphics[width=0.22\textwidth]{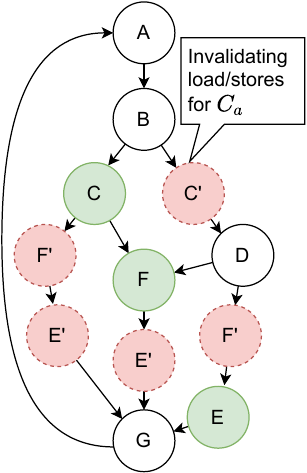}\label{fig:generalSpeculationMain}}%

\caption{A visualization of our CFG transformations to enable speculative LSQ address allocations. Left: address generating CFG; right: compute CFG.}

\label{fig:HoistCFG}
\end{figure}

We now formalize and generalize the transformation from the previous section.
Assume a single LSQ connected to an address generating CFG and compute CFG.
The ordering of speculative allocations across loops is trivial, as is ordering of speculative allocations relative to non-speculative ones (blocks that share a control-dependency source will all have either speculative or non-speculative allocations).
The key question that we answer is how to preserve the relative order between speculative address allocations made in the address generating CFG and the invalidating load/stores in the compute CFG.

\vspace{0.5em}
\textit{Definition 1:} Let a \textit{special control-dependency} relationship be a control-dependency between basic blocks $A$ and $B$ (written $A \xrightarrow{scd} B$), such that $A$ is control-dependent on $B$, $B$ is not a loop header, and the branch in $B$ depends on values loaded from an LSQ. 
If $A \xrightarrow{scd} B$, then we say that $B$ is the special control-dependency source block of $A$.
\vspace{0.5em}

\textit{Definition 2:} Let $\mathbb{B}$ be the set of basic blocks with memory operations selected to be routed through an LSQ, such that each $B \in \mathbb{B}$ has a special control dependency.

\vspace{0.5em}
\textit{Definition 3:} Let $B_{a}$ be the set of all address allocations for a given block $B \in \mathbb{B}$.
Since the special control-dependency relation applies to whole blocks, all $B_{a}$ allocations are speculative. Ordering of invalidations within a block is trivial.

\vspace{0.5em}
\textit{Definition 4:} Let a \textit{poison basic block} $B'$ be a basic block which invalidates misspeculated LSQ address allocations $B_a$ corresponding to memory operations in block $B$. 
Each CFG path should contain either $B$ or $B'$, but never both (a speculated address allocation should either be used or invalidated on every path, but never both).

\vspace{0.5em}
\textit{Definition 5:} Block $B$ becomes unreachable when the CFG $edge = (E_{start}, E_{end}$) is taken if there exists a CFG path from $E_{start}$ to $B$ but not from $E_{end}$ to $B$. 
Only paths within a loop are considered, loop back edges constitute the end of a path.
\vspace{0.5em}

\textit{In the address generating CFG}: we iteratively move up the address allocations $B_{a}$ for every $B \in \mathbb{B}$ to the end of its special control-dependency source block.
If a $B \in \mathbb{B}$ block has multiple such source blocks, then we pick one at random.
Every special control-dependency source block keeps a stack of address allocations moved to it. 
We first push on the stack allocations moved from the left sub-graph, then the right (the choice between left and right is arbitrary but has to be consistent).
When there are no more basic blocks with LSQ allocations that have a special control-dependency, then we stop.
At this point, each block in the CFG that contains speculative allocations will also have a stack exactly representing the order of these allocations.

\textit{Example:} fig. \ref{fig:generalSpeculationAddress} shows an address gemerating CFG with basic blocks $C$, $F$, and $E$ containing LSQ address allocations, and $F \xrightarrow{scd} C$, $F \xrightarrow{scd} D$, $E \xrightarrow{scd} D$, $C \xrightarrow{scd} B$.
There will be two iterations of hoisting.
On the first iteration, $F_a$ moves to $C$, $E_a$ moves to $D$, and $C_a$ moves to $B$.
On the second iterations, $F_a$ (now in block $C$) moves to $B$, and $E_a$ (now in block $D$) moves to $B$.
Block $B$ has no special control-dependency, so we stop after two iterations.
Block $B$ will have the following speculative allocation stack: $\{C_a, F_a, E_a\}$.

\algrenewcommand\algorithmicindent{1em}%

\begin{algorithm}[t]
\caption{Insertion Of Poison Basic Blocks}\label{alg:speculation}
\begin{algorithmic}
\State \textbf{Input: } loop CFG; basic block $B_{spec}$; allocation stack $L$

\For{$B_{a} \in L$}
\For{$edge \in \text{cfg\_traversal}(B_{spec})$}

    \State $C_1 \gets$ $B$ becomes unreachable when $edge$ is taken
    
    \State $C_2 \gets$ no $D_a \in L$ s.t. $D$ is reachable from 
    \State \hspace{25pt} $edge$ and $D_a$ precedes $B_a$ in $L$
    
    
    \If {$C_1$ \textbf{and} $C_2$}
        \State create poison block $B'$ on $edge$
    \EndIf

\EndFor
\EndFor

\end{algorithmic}
\end{algorithm}


\textit{In the compute CFG}: we insert new basic blocks with invalidating LSQ loads/stores on CFG edges where the memory operation corresponding to a given speculated address allocation becomes unreachable.
Alg. \ref{alg:speculation} presents pseudocode for our insertion procedure.
It takes as input a basic block $B_{spec}$ and stack $L$. 
Stack $L$ contains ordered speculative address allocations hoisted to $B_{spec}$ (the result of the hoisting procedure from the previous paragraph).
\vspace{0.5em}

\textit{Theorem.} Alg. \ref{alg:speculation} transforms the compute CFG such that on every CFG path starting at $B_{spec}$ each speculated address allocation $B_a \in L$ is either used or invalidated, but never both. And the relative order of uses or invalidations matches the order of speculated allocations in the address generating CFG, i.e. in stack $L$ order.

\textit{Proof.} The proof follows from the construction of alg. \ref{alg:speculation}.
The algorithm goes over all $B_{a} \in L$ in their allocation order.
For each such $B_a$, it visits every CFG edge dominated by $B_{spec}$ in control-flow order.
At each edge, an invalidating $B'$ will only be inserted if taking that edge will make block $B$ unreachable (condition $C_1$), and if preceding allocations in $L$ have already been used or invalidated (condition $C_2$).
Thus, on every CFG path starting at $B_{spec}$, each $B_a \in L$ will have either been used or invalidated (but not both) in the $L$ allocation order.
\hfill $\blacksquare$
\vspace{0.5em}

\textit{Example:} fig. \ref{fig:generalSpeculationMain} shows how poison blocks would be inserted given the address generation loop from fig. \ref{fig:generalSpeculationAddress}.
Note how $E'$ is not inserted on the $(B, C)$ edge because of condition $C_2$ in the algorithm: $C_a$ precedes $E_a$ in $L$ and is still reachable from the $(B, C)$ edge.

The transformation in the compute CFG has no misspeculation overhead.
Any superfluously created basic block will be removed using existing CFG simplification algorithms.
After simplification, the example CFG from fig. \ref{fig:generalSpeculationMain} would only have two, not five, new basic blocks (one on the $(C, G)$ edge, and one on the $(D, E)$ edge).
For some speculation scenarios there need not be any poison blocks, e.g., in if-then-else branches, where each branch contains the same memory operations using the same address expressions.

\subsection{Optimal Store Allocation Queue Size} \label{sec:lsqSizing}

The optimal size of our store allocation queue depends on the target loop initiation interval (II).
Assume a target II of 1, and a loop datapath as presented in our motivating example in fig. \ref{fig:MotivatingExampleCode}.
Assume \texttt{f(x)} has a latency of $L$ and that there are no true data hazards, so an actual II of 1 is possible at runtime.
To achieve this II, at iteration $N$ our LSQ should be able to disambiguate a load address for iteration $N+L$.
This requires the LSQ to be able to hold $L$ store allocations to cover all store addresses for the $[N, N+L]$ iteration range.
Thus, the optimal store allocation queue size in this case is equal to the maximum latency between a dependent load and a store, call this $maxLoadToStoreDelay$ (for most codes, this is equal to the recurrence constrained II discussed in the introduction).
The optimal size will increase if there are multiple stores in the loop datapath, call this number $numStoresInLoop$.
All of the above information is static, allowing us to find an optimal store allocation queue size at compile time:

$$\ceil[\Big]{\frac{maxLoadToStoreDelay}{targetII} \times numStoresInLoop}$$
\vspace{0.5em}

Tab. \ref{table:scalabilityTableQueue} shows how the resource usage and critical path of our LSQ scales with the size of the store allocation queue.

\section{Evaluation and Results} \label{sec:evaluation}

In this section, we evaluate our work against two commercial HLS compilers (Intel HLS \cite{intelHLS} and Vivado HLS \cite{vivado}), and against a dynamically scheduled academic HLS compiler that uses a state of the art LSQ \cite{josipovic_dynamatic_2022}.
We also show how our LSQ design scales with the size of its store allocation queue.

\subsection{Methodology}

Our compiler analysis' and transformations are implemented as LLVM passes, and we integrated them with the Intel SYCL HLS compiler \cite{sycl_github}.
We automatically find data hazards in the input code, decouple the address generation into separate modulo-scheduled pipelines (separate SYCL kernels), and connect memory requests to an LSQ specialized to the input code.
Our compiler passes and LSQ are publicly available\footnote{\url{https://github.com/robertszafa/elastic-sycl-hls}}.

We evaluate our work against the dynamic HLS tool Dynamatic using a research artifact from their most recent paper \cite{straight_to_the_queue}.
Cycle counts were obtained using ModelSim and are compared directly between all tools.
Dynamatic uses Vivado for synthesis, while we use Intel tools, making a direct comparison of area and circuit frequency in absolute terms difficult.
Instead, we compare the \textit{normalized} frequency, execution time, and area overhead of Dynamatic and our approach against their respective static HLS baseline.
For Dynamatic we used Vivado 2019.2 and the Xilinx xc7k160tfbg484 FPGA.
For our approach, we used Quartus 19.2 and the Altera 10AX115S FPGA.
All benchmarks are integer based to avoid differences in floating point performance across FPGAs.
When comparing against Dynamatic, we only consider codes using on-chip BRAM (we could not use DRAM in Dynamatic).

\renewcommand\theadalign{bc}
\renewcommand\theadfont{\bfseries}
\renewcommand\theadgape{\Gape[4pt]}
\renewcommand\cellgape{\Gape[4pt]}

\begin{table*}[!t]
\centering
\begin{threeparttable}

\renewcommand{\arraystretch}{1.2}
\setlength{\tabcolsep}{1.7pt}
\caption{A comparison of our work against Vivado, Dynamatic \cite{straight_to_the_queue}, and Intel HLS. All codes use on-chip BRAM.}
\label{table:benchmarkTable}
\centering
\begin{tabular}{l | cccc | cccc | cccccc | cccccc }
\hline
\multirow{2}{*}{\textbf{Benchmark}} & \multicolumn{4}{c|}{\textbf{Cycles (thousands)}} & \multicolumn{4}{c|}{\textbf{Freq. (MHz)}} &  \multicolumn{6}{c|}{\textbf{Execution time ($\mu$s)}}  &  \multicolumn{6}{c}{\textbf{Area (Slices / ALMs)}} \\
& V & D & I & \multicolumn{1}{c|}{O} 
& V & D & I & \multicolumn{1}{c|}{O} 
& V & D & D$/$V & I & O & \multicolumn{1}{c|}{O$/$I}
& V & D & D$/$V & I & O & \multicolumn{1}{c}{O$/$I} \\
\hline
\hline

histogram & 2 & 1--3 & 2.1 & 1-2 & 379 & 155 & 379 & 337 & 5.3 & 6.5-19.4 & \textbf{1.23--3.68} & 5.5 & 3--6 & \textbf{0.55--1.1} & 129 & 5582 & \textbf{43.3} & 1814 & 9847 & \textbf{5.4} \\

getTanh & 68 & 2.5--79 & 56.2 & 1.1--59 & 266 & 89 & 377 & 346 & 263 & 28.1--888 & \textbf{0.11--3.47} & 149 & 4.1--224 & \textbf{0.03--1.51} & 572 & 22399 & \textbf{39.2} & 1825 & 12753 & \textbf{7} \\

getTanhDouble & 14 & 1--19 & 13.1 & 1--17 & 304 & 96 & 330 & 297 & 46.1 & 10.7--198 & \textbf{0.23--4.3} & 39.8 & 3.5--57.3 & \textbf{0.09--1.44} & 245 & 22103 & \textbf{90.2} & 3803 & 16730 &	\textbf{4.4} \\

vecTrans & 30 & 1.5--31 & 30.1 & 1.1--33 & 304 & 97 & 365 & 291 & 98.7 & 15.9--320 & \textbf{0.16--3.24} & 82.5 & 3.6--113 & \textbf{0.04--1.38} & 125 & 22997 & \textbf{184} & 1811 & 11672 & \textbf{6.4} \\

spmv & 2.3 & 0.8--2.7 & 3.6 & 0.8--2.7 & 263 & 152 & 328 & 280 & 8.7 & 5.2--17.6 & \textbf{0.6--2.02} & 10.9 & 3--9.8 & \textbf{0.28--0.9} & 494 & 5628 & \textbf{11.4} & 5255 & 23406 & \textbf{4.5} \\

chaosNCG & 72 & 37--74 & 74.3 & 2.1--77 & 308 & 155 & 335 & 246 & 234 & 239--477 & \textbf{1.02--2.04} & 222 & 8.4--313 & \textbf{0.04--1.41} & 779 & 2017 & \textbf{2.6} & 5274 & 32960 & \textbf{6.2} \\

BNN & 20 & 15--30 & 20.7 & 10.4--20.4 & 258 & 116 & 365 & 284 & 77.5 & 129--259 & \textbf{1.67--3.34} & 56.9 & 36.8--72 & \textbf{0.65--1.26} & 1214 & 7466 & \textbf{6.2} & 4214 & 20222 & \textbf{4.8} \\

histogramIf & 2 & 5--6 & 2.1 & 1-2.5 & 388 & 117 & 379 & 328 & 5.15 & 42.7-51.3 & \textbf{8.29--8.3} & 5.5 & 3.1--7.7 & \textbf{0.57--1.4} & 155 & 5395 & \textbf{34.8} & 1814	& 10452 & \textbf{5.8} \\

matching & 6 & 6--8 & 7.6 & 2--8.8 & 404 & 110 & 246 & 291 & 14.9 & 54.6--72.7 & \textbf{3.67--4.9} & 30.9 & 7--30.2 & \textbf{0.23--0.98} & 141 & 3778 & \textbf{26.8} & 7713 & 18310 & \textbf{2.4} \\

floydWarshall & 6.2 & 7--11 & 6.3 & 3.4 & 366 & 90 & 229 & 299 & 16.9 & 77.8--122 & \textbf{4.59--7.2} & 27.3 & 11.3 & \textbf{0.42} & 255 & 2226 & \textbf{8.7} & 807 & 5056 & \textbf{6.3} \\

bitonicSort & 3.1 & 2.6--6.1 & 9.6 & 1.5 & 300 & 97 & 248 & 305 & 10.4 & 26.9--62.8 & \textbf{2.58--6} & 38.8 & 4.8 & \textbf{0.12} & 51 & 5683 & \textbf{111} & 911 & 5424 & \textbf{6} \\

\hline
Harmonic mean & \multicolumn{2}{r}{\textbf{0.15--1.4}} &  \multicolumn{2}{r|}{\textbf{0.07--0.64}}  &  & \textbf{0.35} &  & \textbf{0.89} &  &  & \textbf{0.45--3.67} &  &  & \textbf{0.09--0.62} &  &  & \textbf{12.3} &  &  & \textbf{4.9} \\
\hline


\end{tabular}
\begin{tablenotes}
\item {\begin{tabular}{llll}


V -- Vivado HLS \hspace{4pt} & D -- Dynamatic \hspace{4pt} & I -- Intel HLS \hspace{4pt} & O -- Our work  \\
\end{tabular}}
\end{tablenotes}

\end{threeparttable}
\end{table*}

We applied our approach to eleven benchmarks with data hazards used in previous work \cite{dass, josipovic_dynamatic_2022}.
The codes and evaluation results for all tools are available as a public artifact \cite{szafarczyk_zenodo_fpt}.
The addresses in the first seven benchmarks can be decoupled without speculation:
\begin{enumerate}
    \item \textit{histogram} is the code from fig. \ref{fig:MotivatingExampleCode} (loop II=2).
    \item \textit{getTanh} performs a \textit{tanh(x)} approximation on a sparse array (loop IIs=56, 1, 1). 
    \item \textit{getTanhDouble} is similar but uses only one loop, not three (loop II=13). 
    \item \textit{vecTrans} applies a polynomial expression on elements of a sparse array (loop II=30). 
    \item \textit{spmv} is a sparse matrix-vector multiply (loop IIs=1, 9). 
    \item \textit{chaosNCG} is a function from a chaos engine with data-dependent loads and stores (loop II=74). 
    \item \textit{BNN} is a binarized neural network (loop IIs=1, 2, 2). 
\end{enumerate}
The remaining benchmarks have control-dependent stores, making our speculative address allocation approach applicable:
\begin{enumerate}
  \setcounter{enumi}{7}
    \item \textit{histogramIf} is similar to \textit{histogram}, but the store is control dependent on the load value (loop II=2). 
    \item \textit{matching} is the code example from fig. \ref{fig:mmOriginal} (loop II=7). 
    \item \textit{floydMarshall} finds shortest paths in a weighted digraph (loop IIs=1, 1, 6). 
    \item \textit{bitonicSort} sorts a list of integers using a bitonic merge network (loop IIs=1, 1, 7). 
\end{enumerate}
We report worst- and best-case performance, which depends on the true number of data hazards in the input data distribution.
We automatically choose our store allocation queue size according to sec. \ref{sec:lsqSizing}.
For Dynamatic, we manually choose the smallest queue size that enables perfect pipelining in the case of no data hazards, following their approach \cite{liu_lsq_sizing}.

\definecolor{seabornBlue}{RGB}{76,114,176}
\definecolor{seabornGreen}{RGB}{85,168,104}
\definecolor{seabornRed}{RGB}{196,78,82}

\begin{figure}[t!]
\centering

\begin{tikzpicture}
\centering
\begin{axis}[
    x tick label style={font=\footnotesize, color=white},
    y tick label style={font=\footnotesize},
    width=0.475\textwidth,
    height=4cm,
    symbolic x coords={histogram, getTanh, getTanhDouble, vecTrans, spmv, chaosNCG, BNN, histogramIf, matching, floydWarshall, bitonicSort, dummy},
    ylabel={Speedup (log)},
    ymin=-10, ymax=64,
    ymode=log,
    ytick={0, 0.25, 0.5, 1, 2, 4, 8, 16, 32, 64},
    log ticks with fixed point,
	enlargelimits=0.01,
	legend style={at={(0.5,1.3)}, anchor=north,legend columns=-1},
    ymajorgrids=true,
	ybar interval=0.5
]

\addplot[fill=seabornRed, color=seabornRed, opacity=1, error bars/.cd, y dir=minus, y explicit, error bar style={opacity=1, color=seabornRed, mark size=2.5pt, line width=1.5pt, xshift=4.5pt}] coordinates {(histogram,0.81) -= (0,0.54) += (0,0) (getTanh,9.10) -= (0,8.81) += (0,0) (getTanhDouble,4.31) -= (0,4.08) += (0,0) (vecTrans,6.23) -= (0,5.92) += (0,0) (spmv, 1.67) -= (0,1.17) += (0,0) (chaosNCG, 0.98) -= (0,0.49) += (0,0) (BNN,0.6) -= (0,0.3) += (0,0) (histogramIf,0.12) -= (0,0.001) += (0,0) (matching,0.27) -= (0,0.7) += (0,0) (floydWarshall,0.217) -= (0,0.078) += (0,0) (bitonicSort,0.387) -= (0,0.22) += (0,0) (dummy, 1)};

\addplot[fill=seabornBlue, color=seabornBlue, opacity=1, error bars/.cd, y dir=minus, y explicit, error bar style={opacity=1, color=seabornBlue, mark size=2.5pt, line width=1.5pt, xshift=13.3pt}] coordinates {(histogram,1.83) -= (0,0.91) += (0,0) (getTanh,36.37) -= (0,35.71) += (0,0) (getTanhDouble,11.41) -= (0,10.72) += (0,0) (vecTrans,22.83) -= (0,22.1) += (0,0) (spmv, 3.58) -= (0,2.47) += (0,0) (chaosNCG, 26.4) -= (0,25.69) += (0,0) (BNN,1.55) -= (0,0.76) += (0,0) 
(histogramIf,1.76) -= (0,1.04) += (0,0) (matching,4.39) -= (0,3.37) += (0,0) (floydWarshall,2.41) -= (0,0) += (0,0) (bitonicSort,8.1) -= (0,0) += (0,0) (dummy, 1)};

\legend{Dynamatic,This Work}
\end{axis}
\end{tikzpicture}

\vspace{-0.5cm}
\hspace*{-11pt}
\begin{tikzpicture}
\centering
\begin{axis}[
    x tick label style={font=\footnotesize, rotate=40, anchor=east, align=center, yshift=-2pt, xshift=6pt},
    y tick label style={font=\footnotesize},
    width=0.475\textwidth,
    height=4cm,
    symbolic x coords={histogram, getTanh, getTanhDouble, vecTrans, spmv, chaosNCG, BNN, histogramIf, matching, floydWarshall, bitonicSort, dummy},
	ylabel={Area overhead (log)},
    ymin=0, ymax=256,
    ymode=log,
    log ticks with fixed point,
    ytick={1,2,4,8,16, 32, 64, 128, 256},
	enlargelimits=0.01,
	ybar interval=0.5,
    ymajorgrids=true,
]

\addplot[fill=seabornRed, color=seabornRed] coordinates {(histogram,43.3) (getTanh,39.2) (getTanhDouble,90.2) (vecTrans, 184) (spmv, 11.4) (chaosNCG,2.6) (BNN,6.2) (histogramIf, 34.8) (matching,26.8) (floydWarshall, 8.7) (bitonicSort, 111) (dummy, 1)};

\addplot[fill=seabornBlue, color=seabornBlue] coordinates {(histogram,5.4) (getTanh,7) (getTanhDouble,4.4) (vecTrans, 6.4) (spmv, 4.5) (chaosNCG,6.2) (BNN,4.8) (histogramIf, 5.8) (matching,2.4) (floydWarshall, 6.3) (bitonicSort, 6) (dummy, 1)};

\legend{}
\end{axis}
\end{tikzpicture}

\vspace{-0.2cm}

\caption{Speedup and area overhead of our work and Dynamatic \cite{straight_to_the_queue} compared to their static HLS baselines (Intel HLS and Vivado, respectively). The range bars represent the speedup range, with a value below 1 indicating a slowdown.}

\label{fig:AreaAndPerfPlot}

\end{figure}
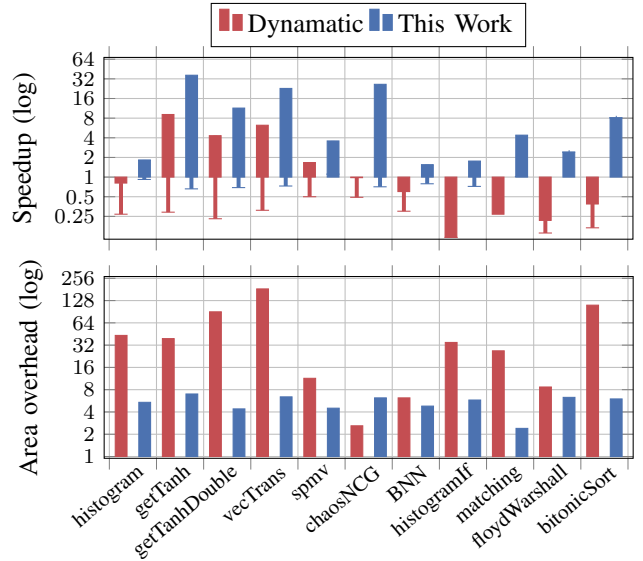

\subsection{BRAM Results}

\textit{Speedup:} Fig. \ref{fig:AreaAndPerfPlot} shows that our approach achieves a higher speedup than Dynamatic when comparing each tool to their respective static HLS baseline.
On most codes, the higher speedup is due to the higher frequency achievable by our LSQ.
On some codes (e.g. \textit{chaosNCG}), we also achieve a better throughput than Dynamatic, because we can support the required large store queue size and Dynamatic cannot (tab. \ref{table:scalabilityTableQueue}).

Tab. \ref{table:benchmarkTable} shows detailed benchmark results.
On average, designs with our LSQ achieve 89\% of the frequency achieved by Intel HLS, whereas Dynamatic LSQ designs achieve a frequency of 35\% compared to Vivado.
Dynamatic sees a higher throughput overhead when the data distribution favors static scheduling (more iterations with a true data hazard), resulting in an average 1.4$\times$ more cycles to finish than the Vivado designs, rising to 3.67$\times$ more execution time due to their lower frequency.
On average, our approach has no overhead in execution time compared to its Intel HLS baseline.
The slight overhead in the number of cycles for some codes is only for data distributions that repeatedly read and write to the same memory location, which is a highly unlikely scenario.

The last four codes benefit from our speculative address allocation scheme, allowing for speedups compared to all other evaluated tools.
The dataflow circuits produced by Dynamatic suffer a loss of address generation decoupling and don't result in any throughput improvements compared to a static pipeline.
Our speculation approach doesn't suffer from loss of decoupling, allowing for improved pipelining.
The results confirm that our approach doesn't see misspeculation overhead.

\textit{Area overhead:} In addition to a better speedup, our LSQ also has a lower area overhead than Dynamatic.
On average, we see a 4.9$\times$ area overhead compared to 12.3$\times$ for Dynamatic, and that is despite the fact that for several codes we use a larger queue size.
We expect the area overhead to be negligible for larger codes, even when using large store queues.

\renewcommand\theadalign{bc}
\renewcommand\theadfont{\bfseries}
\renewcommand\theadgape{\Gape[4pt]}
\renewcommand\cellgape{\Gape[4pt]}

\begin{table}[!t]
\centering
\renewcommand{\arraystretch}{1.2}
\setlength{\tabcolsep}{5pt}


\centering
\caption{Scalability of our store allocation queue compared to the store queue in Dynamatic \cite{straight_to_the_queue} on the histogram benchmark.}
\label{table:scalabilityTableQueue}
\begin{tabular}{c || cc|cc || cc|cc }
\hline
\multirow{2}{*}{\shortstack{\textbf{Queue}\\\textbf{Size}}} & \multicolumn{4}{c||}{\textbf{Freq (MHz)}} & \multicolumn{4}{c}{\textbf{Area (Slices / ALMs)}} \\
& Dyn & \multicolumn{1}{c}{$\times$} & Ours & \multicolumn{1}{c||}{$\times$} & Dyn & \multicolumn{1}{c}{$\times$} & Ours & $\times$ \\

\hline
\hline
No LSQ & 379 & \textbf{1} & 379 & \textbf{1} & 129 & \textbf{1} & 1814 & \textbf{1} \\
2 & 173 & \textbf{0.46} & 338 & \textbf{0.89} & 409 & \textbf{3.2} & 9155 & \textbf{5} \\
4 & 178 & \textbf{0.47} & 337 & \textbf{0.89} & 684 & \textbf{5.3} & 9305 & \textbf{5.1} \\
8 & 163 & \textbf{0.43} & 331 & \textbf{0.87} & 1554 & \textbf{12} & 9847 & \textbf{5.4} \\ 
16 & 155 & \textbf{0.41} & 313 & \textbf{0.83} & 5582 & \textbf{43} & 10705 & \textbf{5.9} \\
32 & 92 & \textbf{0.24} & 271 & \textbf{0.72} & 22580 & \textbf{175} & 12509 & \textbf{6.9} \\ 
64 & - & - & 274 & \textbf{0.72} & - & - & 14140 & \textbf{7.8} \\ 
128 & - & - & 258 & \textbf{0.68} & - & - & 23623 & \textbf{13} \\ 
256 & - & - & 195 & \textbf{0.51} & - & - & 39598 & \textbf{22} \\ 


\end{tabular}


\end{table}

\textit{Store queue size scalability:}
Tab. \ref{table:scalabilityTableQueue} shows how the frequency and area usage changes with the size of our store allocation queue.
Previous LSQ designs targeting FPGAs are notorious for their poor scalability \cite{mem_disamb_approaches, liu_lsq_sizing}.
Our LSQ scales better, allowing for store queues with hundreds of entries.
Even though larger store queues still degrade the achievable circuit frequency, the degradation is sub-linear and is more than compensated by the increased potential throughput compared to statically scheduled memory accesses. 
For example, for a 256 entry store queue, the circuit frequency drops by 2$\times$, but the potential throughput increases by 256$\times$ in the best case.

\subsection{DRAM Results}

Tab. \ref{table:benchmarkTableDRAM} shows the speedups over static Intel HLS that are possible when using our LSQ to protect DRAM.
In this experiment, we report execution time when running in hardware on the Intel PAC Arria 10 GX FPGA board using dual-channel DDR4 memory.
On average, using our LSQ results in an 4--10$\times$ speedup over Intel HLS.
The store commit queue, needed to cover the maximum store latency to DRAM, has a cache-like effect which is more noticeable in DRAM codes, compared to codes using BRAM.
As a result, our LSQ still offers a significant speedup even if most of the iterations have a true data hazard.

The DRAM benchmarks achieve on average a 7--10$\times$ lower throughput than the BRAM codes.
Our DRAM load-store units do not take advantage of burst reads or writes that amortize the large off-chip memory latency in typical HLS designs.
It is unlikely that DRAM bursts could be used effectively in an LSQ, because the memory access pattern of codes using LSQs
is seldom contiguous.

Circuits with DRAM connections use more resources, making the area overhead of our LSQ smaller (1.4$\times$ for DRAM vs. 3.4$\times$ for BRAM).
For some codes, using our LSQ results in virtually no resource increase.
This is because the Intel HLS baseline uses more costly bursting DRAM load-store units, while we use simpler, pipelined units.

\renewcommand\theadalign{bc}
\renewcommand\theadfont{\bfseries}
\renewcommand\theadgape{\Gape[4pt]}
\renewcommand\cellgape{\Gape[4pt]}

\begin{table}[!t]
\centering
\begin{threeparttable}

\renewcommand{\arraystretch}{1.2}
\setlength{\tabcolsep}{1.6pt}
\caption{Performance of our LSQ when protecting off-chip DRAM.}
\label{table:benchmarkTableDRAM}
\centering
\begin{tabular}{l | ccc | ccc | ccc }
\hline
\multirow{2}{*}{\textbf{Benchmark}} & \multicolumn{3}{c|}{\textbf{Exec. Time ($\mu$s)}} & \multicolumn{3}{c|}{\textbf{Freq. (MHz)}} & \multicolumn{3}{c}{\textbf{Area (ALMs)}} \\
& I & O & \multicolumn{1}{c|}{O$/$I}
& I & O & \multicolumn{1}{c|}{O$/$I}
& I & O & \multicolumn{1}{c}{O$/$I} \\
\hline
\hline

histogram & 363 & 43.7--61.3 & \textbf{0.12--0.17} & 273 & 272 & \textbf{1} & 19832 & 19647 & \textbf{1} \\
getTanh & 564 & 36.9--150 & \textbf{0.08--0.27} & 281 & 205 & \textbf{0.73} & 27365 & 35559 & \textbf{1.3} \\
getTanhDouble & 396 & 35.8--122 & \textbf{0.09--0.31} & 281 & 235 & \textbf{0.84} & 26018 & 29051 & \textbf{1.1} \\
vecTrans & 441 & 40.6--182 & \textbf{0.09--0.37} & 305 & 241 & \textbf{0.79} & 20217 & 22814 & \textbf{1.1} \\
spmv & 158 & 40.8--63.5 & \textbf{0.26--0.34} & 287 & 256 & \textbf{0.89} & 7826 & 18313 & \textbf{2.3} \\
chaosNCG & 687 & 63.3--502 & \textbf{0.09--0.54} & 270 & 170 & \textbf{0.63} & 21190 & 37314 & \textbf{1.8} \\
BNN & 4167 & 336--636 & \textbf{0.08--0.15} & 264 & 241 & \textbf{0.91} & 10916 & 17459 & \textbf{1.6} \\
histogramIf & 362 & 34.2--61.8 & \textbf{0.09--0.17} & 274 & 248 & \textbf{0.91} & 19903 & 20950 & \textbf{1.1} \\
matching & 496 & 53.5--175 & \textbf{0.11--0.35} & 289 & 227 & \textbf{0.79} & 8655 & 19907 & \textbf{2.3} \\
floydWarshall & 300 & 59.9--98.4 & \textbf{0.21--0.33} & 257 & 250 & \textbf{0.97} & 31280 & 32173 & \textbf{1} \\
bitonicSort & 319 & 33.9--53.3 & \textbf{0.11--0.17} & 270 & 241 & \textbf{0.89} & 12587 & 24781 & \textbf{2} \\

\hline
Harmonic mean &  &  & \textbf{0.1--0.25} &  &  & \textbf{0.84} &  &  & \textbf{1.4} \\
\hline



\end{tabular}
\begin{tablenotes}
\item {\begin{tabular}{ll}
I -- Intel HLS \hspace{4pt} & O -- Our work  \\
\end{tabular}}
\end{tablenotes}

\end{threeparttable}
\end{table}


\section{Conclusion}

We presented a novel, shift-register-based load-store queue (LSQ) design adapted to spatial architectures and tightly coupled with an HLS compiler that can specialize parts of the LSQ to a given target code.
We introduced the concept of speculative address allocations to the LSQ, which enables out-of-order loads on a broader range of codes than before with no misspeculation overhead.
Our LSQ design achieves a higher frequency and lower area overhead compared to previous LSQs used in HLS, resulting in an average speedup of 11$\times$ compared to static HLS and $5\times$ compared to dynamic HLS.
Our LSQ scales to queues with hundreds of entries, and can protect both on-chip and off-chip memory.

\section*{Acknowledgments}
This work was partly supported by the UK EPSRC. We thank Intel for access to FPGAs through the FPGA DevCloud, and the anonymous reviewers for improving this paper.

For the purpose of open access, a Creative Commons Attribution (CC BY) license has been applied to the Author Accepted Manuscript version of this paper.

\newpage

\bibliographystyle{IEEEtran}
\bibliography{IEEEabrv, references}

\end{document}